\documentclass[ aps, pra, reprint, superscriptaddress, floatfix,10pt]{revtex4-2}
\usepackage[separate-uncertainty = true,multi-part-units=single]{siunitx}

\DeclareSIUnit\gauss{G}
\DeclareSIUnit{\u}{u}

\usepackage{bm} % bold vectors
\usepackage[dvipsnames]{xcolor}

\usepackage{graphicx} % Required for inserting images
\usepackage{amsfonts,amsmath,amssymb}
\usepackage{amsthm}%

\usepackage{textcomp}%
\usepackage{booktabs}%
\usepackage{algorithmicx}%
\usepackage{algpseudocode}%
\usepackage{listings}%
\usepackage{nicefrac, mathrsfs}
\usepackage{caption,subcaption}
\usepackage{tikz}
\usepackage[hidelinks]{hyperref}
\usepackage[nameinlink]{cleveref}	% needs to be imported after hyperref
\usepackage[percent]{overpic}   % place subfigure captions
\usepackage{placeins}
\usepackage{microtype}
\Crefname{equation}{Eq.}{Eqs.}
\Crefname{figure}{Fig.}{Figs.}
\Crefname{table}{Tab.}{Tabs.}

\usepackage{lineno}
\AtBeginEnvironment{equation}{\linenomath}
\AtEndEnvironment{equation}{\endlinenomath}
\AtBeginEnvironment{align}{\linenomath}
\AtEndEnvironment{align}{\endlinenomath}
\AtBeginEnvironment{align*}{\linenomath}
\AtEndEnvironment{align*}{\endlinenomath}

\newcommand\Bstrut{\rule[-0.9ex]{0pt}{0pt}}   % = `bottom' strut
\newcommand\Tstrut{\rule{0pt}{2.6ex}}         % = `top' strut

\makeatletter
\usepackage{etoolbox}
\patchcmd\H@refstepcounter{\protected@edef}{\protected@xdef}{}{}
\makeatother
\begin{document}

\title{Single-atom billiards in an ultracold bath with energy-tunable collisions and threshold-sensitive detection}
% \title{Few-Collision Friction of Single Atoms in an Ultracold Thermal Bath}
\author{Silvia Hiebel}
\affiliation{
    RPTU University Kaiserslautern-Landau,
    Department of Physics and State Research Center OPTIMAS,
    Kaiserslautern, Germany
}

\author{Sabrina Burgardt}
\affiliation{
    RPTU University Kaiserslautern-Landau,
    Department of Physics and State Research Center OPTIMAS,
    Kaiserslautern, Germany
}

\author{Julian Feß}
\affiliation{
    RPTU University Kaiserslautern-Landau,
    Department of Physics and State Research Center OPTIMAS,
    Kaiserslautern, Germany
}

\author{Artur Widera}
\email{widera@rptu.de}
\affiliation{
    RPTU University Kaiserslautern-Landau,
    Department of Physics and State Research Center OPTIMAS,
    Kaiserslautern, Germany
}

    % \section*{Abstract}
    \begin{abstract}
        Transport processes at the microscopic scale can differ strongly from their macroscopic counterparts, particularly when the environment is not a well-defined surface but an unordered thermal medium. Ultracold quantum gases offer a high degree of controllability and directly observable dynamics, thereby providing a versatile platform to study such transport processes under precisely controllable conditions.
        Here, we realize an energy-tunable single-atom collider by transporting individual cesium atoms at controlled velocity through an ultracold rubidium bath. 
        The constant lattice velocity sets the directed contribution to the collision-energy distribution, while the lattice depth provides a tunable energy threshold for collision-induced transport disruption. 
        By independently varying these parameters together with bath density and temperature, we separate the number of collision events from their probability to disrupt transport.
        Across a regime with only few expected collisions per passage, the inferred single-collision ejection probability rises from small values to near unity as the collision energy exceeds the lattice confinement scale.
        The experiment thus converts microscopic collisional energy transfer into a binary single-atom observable and provides a controlled link between individual scattering events and emergent dissipative transport.
    \end{abstract}

\maketitle

\section{Introduction}\label{introduction}
Whenever only a few collisions occur during transport and each collision strongly affects the transported particle, continuum descriptions of the bath and drag break down, and individual scattering events determine the transport outcome.
Understanding how motion through an environment crosses over from ballistic propagation to dissipative transport is a central problem in fields ranging from trapped-ion nanotribology \cite{Bylinskii.2015, Gangloff.2015} and surface friction \cite{Bhushan.1995} to structural superlubricity on crystals \cite{Giessibl.2002, Hod.2018} and atomic-scale friction-force microscopy \cite{Song.2024}. 
Ultracold quantum gases provide an ideal platform for studying collisional transport in a highly controllable environment.
Concretely, unlike conventional drag experiments, in which dissipation emerges from many unresolved interactions \cite{Krim.2012, Squires.2010}, the system presented here operates in a regime where individual collisions can determine the binary outcome of a transport process: whether an atom is successfully transported to its target region or not.
When both the projectile energy and the confinement threshold are tunable, such a binary outcome can be used as a microscopic probe of collisional energy transfer.
Driven particles in periodic potentials as realized here constitute a paradigmatic model for stochastic transport, Brownian motion, and locked-to-running transitions \cite{Reimann.2002,Haenggi.2009}. 
Optical lattices offer an especially clean realization of such landscapes, but experiments combining a deterministic moving lattice with a real thermal atomic bath at the single-particle level remain rare.
In ultracold degenerate gases, the motion of an impurity is directly connected to the Landau criterion and to the onset of dissipation via the creation of elementary excitations \cite{Raman.1999,Onofrio.2000, Astrakharchik.2004,Wouters.2010}. 
This underlies impurity and polaron physics \cite{Massignan.2014,Massignan.2026}, in which a few atoms spectroscopically probe the response of a many-body environment. 
While most experiments have focused on equilibrium properties and polaron formation, controlled transport of an individual impurity through a bath has only recently become a focus of research \cite{Dieterle.2021,hennebichler.2026}.

Here, we turn individual Cesium-133 (Cs) atoms into controlled microscopic projectiles and transport them through an ultracold Rubidium-87 (Rb) target. 
The constant lattice velocity through the bath sets the directed relative velocity and thereby tunes the collision-energy distribution, whereas the lattice depth sets the energy scale required to disrupt transport.
Independently, the bath density controls the expected number of collision events, while the bath temperature sets the thermal broadening of the relative-velocity distribution.
This enables access to the rarely explored regime in which a single neutral impurity undergoes only a few collisions during transport and dissipation arises from discrete binary scattering events rather than collective processes.
The transported atom acts as a microscopic probe of the bath: its probability of traversing the gas reflects both the local collision rate and the energy transferred in individual scattering events \cite{Spethmann.2012,Hohmann.2017,Schmidt.2019,Adam.2022}. 

Whereas previous work studied the free relaxation and diffusion of individual Cs tracer atoms in a dilute Rb gas \cite{Spethmann.2012,Hohmann.2017}, the present experiment adds a controlled moving lattice potential and asks whether an impurity can be transported successfully through the bath. 
In this way, the lattice converts collisional energy transfer into a binary transport outcome.
The experiment is thus conceptually related to active microrheology, where the response of a medium is inferred from the forced motion of an embedded probe particle \cite{Squires.2010,Gazuz.2009,peng.2022}. Here, the probe is reduced to a single atom and the observable to a binary transport outcome.

\section{Experimental setup}\label{exp_methods}
    We investigate transport processes in ultracold atomic systems by preparing a mixture consisting of an ultracold Rb bath and individual Cs impurity atoms confined in a shared dipole trap. We transport the Cs atoms through the Rb cloud using an optical lattice. Controlled displacement of the lattice potential enables high-precision studies of transport efficiency and impurity–bath dynamics.
    For these experiments, thermal clouds of Rb atoms are prepared at temperatures in the range of \mbox{$T=$ \SIrange{500}{1100}{\nano\kelvin}}, with peak densities of the order of $n_\mathrm{peak} = \SIrange{3e12}{1.8e13}{\per\centi\meter\cubed}$, corresponding to total atom numbers of the order of a few thousand to approximately thirty thousand atoms.
    The large separation between the number of Rb bath atoms and Cs impurities (greater than 100:1) motivates treating the Rb cloud as an external reservoir. The mean free path $l$ of Cs atoms in the Rb cloud depends strongly on atom velocity and bath density. It can therefore vary from a few micrometers to several cloud widths, as discussed further in Appendix D. The Rb bath extension $L_\mathrm{bath}$, here expressed as the full width at half maximum, ranges from \SIrange{115}{155}{\micro\meter}.
    Accordingly, the Knudsen number spans $\mathrm{Kn}=l/L_\mathrm{bath}=0.07$--$7.19$, with values of order unity over much of the relevant parameter range. The experiment therefore accesses the crossover from a collision-dominated to a rarefied regime, where the discrete nature of individual collisions becomes increasingly important.

    We initially laser-cool the Rb atoms in a magneto-optical trap (MOT). Subsequently, evaporative cooling is performed in an anisotropic, crossed optical dipole trap (ODT). We prepare the Rb atoms in the magnetically insensitive hyperfine state $F = 1, m_\mathrm{F} = 0$ of the electronic ground state $5S_{1/2}$, where $F$ denotes the total angular momentum and $m_\mathrm{F}$ its projection onto the quantization axis. 
    This state is obtained by applying an optical-pumping sequence followed by a Landau–Zener radio-frequency sweep.
    The temperature, atom number, and density of the Rb cloud are characterized by time-of-flight absorption imaging.
    Independently, we prepare approximately 20 individual Cs atoms at a temperature of $T_\mathrm{Cs} = \SI{4.0+-0.5}{\micro \kelvin}$. 
    The Cs impurities remain sufficiently dilute to behave as independent single-particle projectiles and are detected individually.
    We first confine and cool them in a MOT, then use degenerate Raman sideband cooling to reduce their temperature further and transfer them into the energetically lowest Zeeman substate, $F = 3, m_\mathrm{F} = 3$. 
    During this cooling stage, the Cs atoms are held in a separate crossed optical dipole trap, spatially separated by \SI{208+-4}{\micro\meter} from the center of the crossed ODT used for the Rb cloud.
    The two atomic species share a common dipole trap that provides horizontal confinement.
    
\begin{figure*}[hbt]
        \begin{subfigure}[t]{0.43\textwidth}
              \caption{}
              \includegraphics{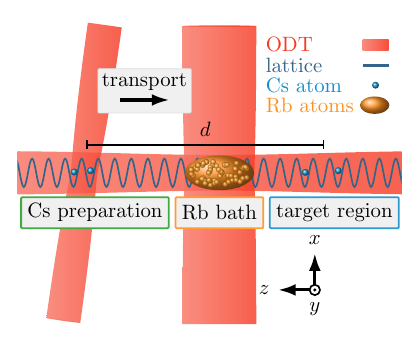}
              \label{fig:exp_setup}
            \end{subfigure}
            \hfill
            \begin{subfigure}[t]{0.54\textwidth}
              \caption{}
              \includegraphics{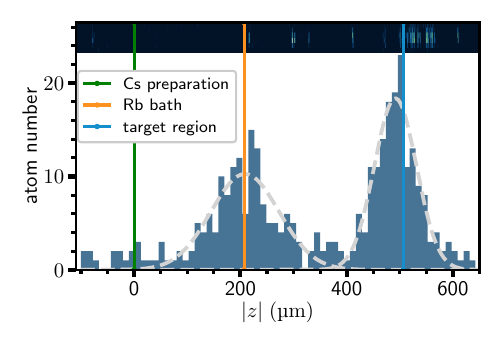}
              \label{fig:histogram}
            \end{subfigure}
	   \caption[Sketch of the experimental setup]{Experimental setup and data evaluation. a) Sketch of the experimental setup. The Rb cloud and the individual Cs atoms are initially prepared in spatially separated regions. Both species are confined radially by a common dipole-trap beam propagating along the $z$ direction. Following the preparation stage, the Cs atoms are loaded into a one-dimensional optical lattice, which enables their controlled transport along the shared dipole trap into or beyond the Rb cloud. b) Exemplary histogram showing the summed Cs atom positions along the $z$ direction for 12 individual measurement runs. The colored vertical lines indicate the centers of the corresponding regions: green line: Cs preparation region, orange line: Rb bath, blue line: target position for the Cs transport. The dashed gray curves are guides to the eye highlighting the atoms in the Rb and the target region. The inset shows one representative fluorescence image projected on the transport axis. 
       }\label{fig:exp_setup_overview}
    \end{figure*}

    After preparing both species, the Cs atoms are transferred from the vertical dipole-trap beam into a one-dimensional lattice potential along the shared dipole trap. The lattice potential is given by
    \begin{equation}
	U(z)=\frac{U_\mathrm{lat}}{2}\cos\left(\frac{4\pi}{\lambda_L}z\right)
    \end{equation}
    with potential depth $U_\mathrm{lat}$ and lattice wavelength $\lambda_L=$ \SI{790.0185}{\nano\meter}. A schematic of the trap setup is shown in \cref{fig:exp_setup}.
    \noindent Two counterpropagating beams with a waist of \SI{29.3}{\micro\meter} create the lattice potential. To minimize the 1D lattice's influence on the Rb bath, we operate at the rubidium tune-out wavelength, where the scalar AC Stark shifts of the Rb D1 and D2 lines cancel.
    For Rb atoms, the polarizability vanishes, and they experience a zero crossing of the total energy shift, thereby minimizing the potential's influence on them \cite{LeBlanc.2007, Leonard.2015, Lamporesi.2010}.
    In contrast, Cs atoms experience a strong repulsive lattice potential and can be trapped in the potential minima. 
    The selectivity of the potential $V_\mathrm{Cs}/V_\mathrm{Rb}$ is about $\SI{3.3e3}{}$. 
    For a more detailed description, see \cite{Schmidt.2016}.
    This strong species selectivity is essential for the collider geometry, as it allows us to accelerate and confine the Cs projectile while leaving the Rb target largely unaffected by the transport potential.
    To transport the Cs atoms along the horizontal dipole-trap beam, we slightly detune one of the lattice beams to move the optical lattice. In our experiments, we target transport distances of $d\approx\SI{500}{\micro\meter}$.
    The resulting moving potential, which is also tilted in the accelerated case, is given by
    \begin{equation}\label{eq:tilted_lattice}
    	U(z)=\frac{U_\mathrm{lat}}{2}\cos\left(\frac{4\pi}{\lambda_L}z\right) + m_\mathrm{Cs} a z
    \end{equation}
    with mass $m_\mathrm{Cs}$ of the Cs atoms and acceleration $a$. \Cref{eq:tilted_lattice} shows that, when the acceleration-induced tilt becomes sufficiently large relative to the lattice depth, the barrier between adjacent lattice sites vanishes and the atoms are no longer confined to individual minima.
    We choose the acceleration such that the Cs atoms reach a well-defined horizontal velocity before entering the Rb cloud.
    The acceleration is then set to zero, and the Cs atoms traverse the cloud at constant velocity. 
    This constant-velocity interaction window is crucial for the collider operation: acceleration is completed before the atom enters the bath, so that the directed component of the relative collision velocity remains well defined during the interaction.\\
    We can compare the thermal energy of the Cs atoms $E_\mathrm{th}=3/2 k_\mathrm{B}T_\mathrm{Cs}$ with the kinetic energy \mbox{$E_\mathrm{lat}=1/2m_\mathrm{Cs}v^2_\mathrm{lat}$} they gain through transport in the lattice at lattice velocity $v_\mathrm{lat}$ via \cite{Schwabl.2006}
    \begin{equation}
        \frac{E_\mathrm{lat}}{E_\mathrm{th}}=\frac{m_\mathrm{Cs} v_\mathrm{lat}^2}{3k_\mathrm{B}T_\mathrm{Cs}}.
        \label{eq:MB_distribution}
    \end{equation}
    At \SI{4}{\micro\kelvin}, this yields ratios of 2.1 for the lowest lattice velocity in \cref{fig:through_colormeshes} and 43.2 for the highest. 
    Thus, over a substantial part of the accessible parameter range, the directed projectile velocity exceeds the thermal velocity spread, enabling controlled tuning of the collision-energy distribution.
    The Maxwell–Boltzmann distribution of a single velocity component is Gaussian with distribution function \cite{pathria2011statistical}
    \begin{equation}
        f(v_z)=\sqrt{\frac{m_\mathrm{Cs}}{2\pi k_\mathrm{B}T_\mathrm{Cs}}}\exp\left(-\frac{m_\mathrm{Cs}v_z^2}{2k_\mathrm{B}T} \right).
    \end{equation} 
    The standard deviation is given by $\sigma_{v,z}=\sqrt{\frac{k_\mathrm{B}T}{m_\mathrm{Cs}}}$. By comparing the width of this velocity distribution and the lattice velocity (${v_\text{lat}}/{\sigma_{v,z}}$), we obtain ratios of 2.5 (11.4) for the lowest (highest) lattice velocity.
    After transporting the Cs atoms, we image them via fluorescence imaging. The resolution of our imaging system limits the accuracy of the position detection to \SI{1.4}{\micro\meter}; therefore, individual lattice sites cannot be resolved.
    \Cref{fig:histogram} illustrates the spatial distribution of the Cs atoms along the $z$-axis. 
    The histogram is the result of the sum of 12 measurements for $v_\mathrm{lat}\approx\SI{10}{\centi\meter\per\second}$ and $U_\mathrm{lat}/k_\mathrm{B} = \SI{6.3}{\micro\kelvin}$ and bath parameters of $n_\mathrm{peak} = \SI{1.15e13}{\per\centi\meter\cubed}$ and $T= \SI{918}{\nano\kelvin}$. The atoms are initially prepared in the Cs preparation region and then transported through the Rb bath toward the target region. If the atom position is detected within the target region, the atom is counted as successfully transported, whereas atoms detected within the Rb bath indicate incomplete transport and therefore increased or at least prolonged interaction with the Rb bath. The transport efficiency is calculated as the fraction of transported atoms divided by the initial Cs atom number.
    For this parameter set, a substantial fraction of atoms, approximately $50\%$ of the initial atom number, remains localized within the Rb bath region. In contrast, approximately $40\%$ of the atoms are successfully transported to the target region. Roughly $5\%$ of the atoms remain in the preparation region, and the remaining fraction is lost during the measurement sequence, where the dominant loss mechanism is three-body recombination in the Rb bath. This lost fraction corresponds to approximately one atom per experimental realization and lies within the range of shot-to-shot fluctuations, such that it does not significantly affect the transport efficiency.

\section{Cs transport measurements}\label{sec:meas_cs}
\begin{figure*}
            \begin{subfigure}[t]{0.4\textwidth}
            	\vspace{-4pt}
              \caption{}
              \includegraphics{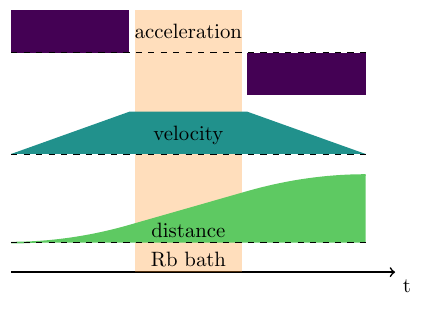}
              \label{fig:acc_vel_dis_scheme}
            \end{subfigure}
            \hfill
            \begin{subfigure}[t]{0.55\textwidth}
            	\vspace{-4pt}
              \caption{}
              \includegraphics{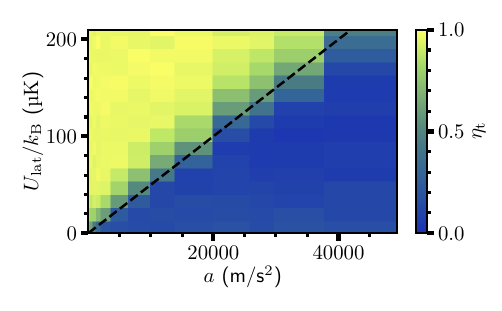}
              \label{fig:transport_efficiency_Cs_2d}
            \end{subfigure}
    	\caption[Cs transport measurements]{Cs atom transport in the one-dimensional lattice. a) Lattice motion during transport. The purple graph shows the acceleration during transport, the teal graph shows the velocity, and the light green curve shows the corresponding distance traveled. The orange background indicates when the atoms traverse the Rb bath.
        b) Transport efficiency of the Cs atoms without additional Rb bath as a function of lattice depth and acceleration. The black dashed line indicates $U_\mathrm{lat}/k_\mathrm{B}=sa$, with $s=\SI{5e{-3}}{\micro\kelvin\second\squared\per\meter}$ and shows the boundary between locked and running behavior.}\label{fig:fast_cs_transport}
    \end{figure*}

    The transport efficiency of single atoms in the optical lattice strongly depends on the transport parameters.
    The simplest transport protocol in our setup is to prepare the Cs atoms, load them into the one-dimensional transport lattice, and apply an acceleration to the lattice, similarly to reference \cite{Matthies.2024}. We use the trapezoidal velocity profile shown in \cref{fig:acc_vel_dis_scheme}: the lattice is first accelerated, then moved at constant velocity through the region occupied by the Rb bath, and finally decelerated. The transport scheme is discussed in detail in Appendix A.
    Here, we use this scheme to evaluate transport efficiencies throughout the entire technically accessible parameter range.
    This allows us to extract the transport efficiencies $\eta_t$ for different lattice depths $U_\mathrm{lat}$ and accelerations $a$ in the absence of further external influences from the Rb bath. 
    Our measurement results are shown in \cref{fig:transport_efficiency_Cs_2d}.
    We use the full parameter range of our lattice potential, with lattice depths up to \SI{210}{\micro\kelvin} (given in units of temperature) and accelerations up to $a\approx$ \SI{43500}{\meter\per\second\squared}. %43,451 \\
    For the highest possible lattice accelerations, even the deepest lattice we can achieve is not sufficient to transport atoms reliably.
    The characteristic region with sufficiently high transport fidelity is shown in \cref{fig:transport_efficiency_Cs_2d}. The linear dependence of the critical lattice depth on acceleration is indicated by the dashed black line, described by the function \( f(a) = s \cdot a \), which divides the plotted parameter space into two distinct regimes. 
    Here, $s$ denotes a proportionality factor for relating the acceleration to a corresponding critical lattice depth required for successful transport.
    In the lower-right regime of the plot, almost no atoms are transported; instead, they remain in the preparation region because the potential barrier $\Delta U$ between neighboring minima and maxima is small compared to the atomic energy. In the upper-left regime, almost all atoms are successfully transported.
    Between these regions, the system undergoes a transition from weak to strong external forces \cite{Iida.2025}, corresponding to biased (Brownian) motion under a constant external force. For small accelerations, the atoms are in a locked state. When the acceleration-induced tilt becomes sufficiently large relative to the lattice depth, the Cs atoms are in a running state and stay in their starting region \cite{Lindner.2016}. The linear dependence on acceleration is consistent with the observations in \cite{Klostermann.2022}.

\section{Transport measurements through a thermal bath}
\label{meas_through}
    We determine the transport efficiencies of single Cs atoms in the optical lattice through a thermal bath of Rb atoms.
    For these measurements, we position the Rb cloud between the Cs preparation region and the transport target region. In this configuration, the Cs atoms are conveyed at constant velocity through the Rb bath by the optical lattice.
    The setup therefore provides three largely independent control knobs for microscopic collisions: $v_{\rm lat}$ tunes their characteristic energy, $n_{\rm Rb}$ tunes their number, and $U_{\rm lat}$ sets the threshold for detecting their effect on the transported atom. The bath temperature additionally controls the thermal broadening of the collision-energy distribution.
    Experimentally, we vary these parameters within the regime where the lattice does not measurably perturb the Rb bath.
    Collisions between the Cs atoms and individual atoms of the Rb cloud govern this interaction.
    Increasing the Rb density increases the probability that a Cs atom undergoes a collision while traversing the bath and therefore increases the probability of incomplete transport. Increasing the transport velocity shortens the interaction time and generally reduces the mean number of collisions.
    As discussed in Appendix C, the lattice velocity has a negligible influence on the Cs atoms themselves during transport. The additional reduction in transport efficiency can therefore be attributed to interactions with the Rb bath.\par
\begin{figure*}[htbp]
    \begin{minipage}[t]{0.49\textwidth}
    \vspace{0pt}
            \centering
            \begin{overpic}[width=\linewidth]%
                {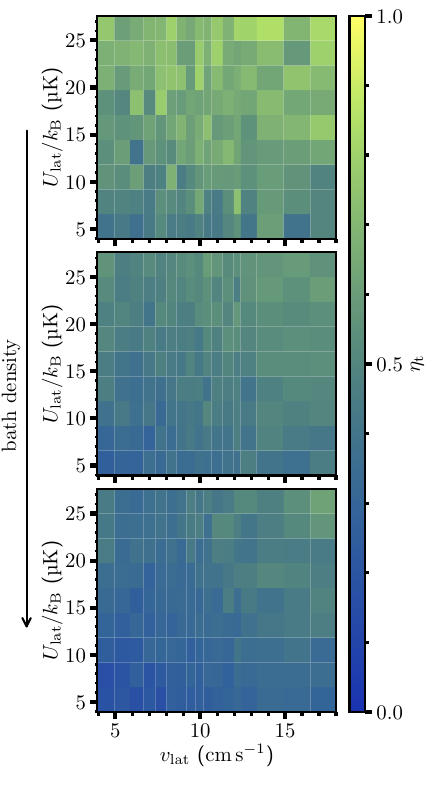}
                \put(13,95){%
                    \fboxsep=1pt%
                    \colorbox{gray!25}{\textbf{a)}}%
                }
                \put(13,65){%
                    \fboxsep=1pt%
                    \colorbox{gray!25}{\textbf{b)}}%
                }
                \put(13,35){%
                    \fboxsep=1pt%
                    \colorbox{gray!25}{\textbf{c)}}%
                }
            \end{overpic}
            \phantomsubcaption
            \label{fig:transport-density-955}
            \phantomsubcaption
            \label{fig:transport-density-918}
            \phantomsubcaption
            \label{fig:transport-density-906}
        \end{minipage}
        \hfill
        %------------------------------------------------
        % Right column: panels (d)--(f)
        %------------------------------------------------
        \begin{minipage}[t]{0.48\textwidth}
        \vspace{6pt}
            \centering
            \begin{overpic}[width=\linewidth]%
                {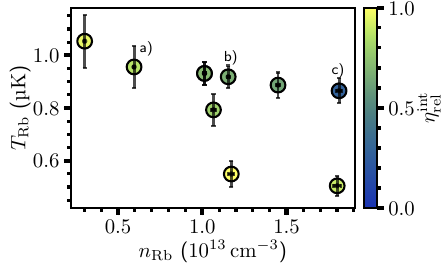}
                \put(0,54){%
                    \fboxsep=1pt%
                    \textbf{d)}%
                }
            \end{overpic}
            \phantomsubcaption
            \label{fig:eta-transp-through}
            \par\vspace{0.0em}
            \begin{overpic}[width=\linewidth]%
                {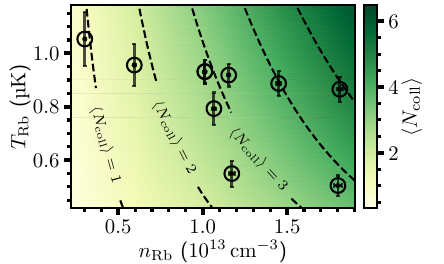}
                \put(0,54){%
                    \fboxsep=1pt%
                    \textbf{e)}%
                }
            \end{overpic}
            \phantomsubcaption
            \label{fig:n-coll-density-temperature-grid}
            \par\vspace{-1.45em}
            \begin{flushleft}
            \begin{overpic}[width=\linewidth]%
                {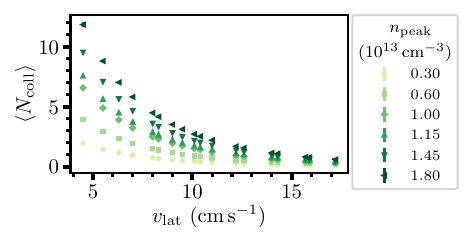}
                \put(0,44){%
                    \fboxsep=1pt%
                    \textbf{f)}%
                }
            \end{overpic}
            \end{flushleft}
            \phantomsubcaption
            \label{fig:ncoll_vs_vlat}
        \end{minipage}
	   \caption[Transport efficiency through thermal bath for exemplary realizations]{Transport efficiency for different bath parameters. a) Low-density: $T=\SI{955+-16}{\nano\kelvin}$, $n=\SI{0.60(7)e13}{\per\centi\meter\cubed}$. b) Intermediate-density: $T=\SI{918+-40}{\nano\kelvin}$, $n=\SI{1.55(6)e13}{\per\centi\meter\cubed}$. c) High-density: $T=\SI{906+-40}{\nano\kelvin}$, $n=\SI{1.80(5)e13}{\per\centi\meter\cubed}$. The displayed results are averages over 15 repetitions for each data point. d) Integrated transport efficiency relative to the efficiency without the Rb bath. The labels indicate the assignment of panels a)-c). e) Maximum average number of collisions for different bath parameters at a velocity of \SI{7}{\centi\meter\per\second}. The black circles indicate the positions of the measurement data from d). The dashed lines indicate contour lines for $\langle N_\mathrm{coll}\rangle = 1-5$. f): Maximum average number of collisions as a function of lattice velocity for a bath temperature of $T\approx \SI{850}{\nano\kelvin}$ and different bath densities assuming transport through the entire Rb bath.}\label{fig:through_colormeshes}
    \end{figure*}

    We show the transport efficiencies of the Cs atoms for similar temperatures but different bath densities of $n=\SI{0.60(7)e13}{\per\centi\meter\cubed}, n=\SI{1.55(6)e13}{\per\centi\meter\cubed}$, and $ n=\SI{1.80(5)e13}{\per\centi\meter\cubed}$ in \cref{fig:through_colormeshes} a)-c).%
    In the absence of the Rb bath, the transport efficiency exceeds $60\%$ over the full parameter range shown in Appendix B.
    For sufficiently dilute Rb clouds, the influence of interspecies collisions on the transport process becomes negligible within the experimental uncertainty. In contrast, at the highest Rb densities investigated, the transport efficiency falls below 60\% across the entire range of transport parameters explored.
    The temperature of the Rb bath is an additional key parameter governing the transport efficiency.
    As the bath temperature increases, a decreasing fraction of Cs atoms is successfully transported through the Rb cloud. 
    The temperature dependence originates from the thermal modification of the relative-velocity distribution and, through the energy-dependent elastic cross section, of the averaged collision-rate coefficient $\langle\sigma_{\rm el}v_{\rm rel}\rangle$.\\
    In \cref{fig:eta-transp-through}, we quantitatively compare the transport efficiencies obtained for different bath conditions. To reduce the parameter space and ensure comparable results, we extract a single efficiency parameter for each pair of bath parameters.
    To this end, we define a transport efficiency integrated over the entire lattice parameter range shown in panels a)-c)
    \begin{equation}
        \eta^\mathrm{int}_\mathrm{rel} = \frac{ \int \int \eta_\mathrm{t}(v_\mathrm{lat},U_\mathrm{lat}) \, \mathrm{d}U_\mathrm{lat}\, \mathrm{d}v_\mathrm{lat}}{\int \int \eta_\mathrm{t, Cs}(v_\mathrm{lat},U_\mathrm{lat}) \, \mathrm{d}U_\mathrm{lat}\, \mathrm{d}v_\mathrm{lat}},
    \end{equation}
     where $\int \int \eta_\mathrm{t, Cs}(v_\mathrm{lat},U_\mathrm{lat}) \mathrm{d}U_\mathrm{lat} \mathrm{d}v_\mathrm{lat}$ denotes the integrated transport efficiency of Cs atoms in the absence of the Rb bath. The corresponding transport efficiencies for Cs-only are discussed in more detail in Appendix C, \cref{fig:Cs_transport_small_params}.
     The results of the measurements shown in \cref{fig:through_colormeshes} a)-c) are indicated by the corresponding labels in \cref{fig:eta-transp-through}.\\

    A central contribution to the reduction in transport efficiency through the bath is the scattering between the Cs atoms and the Rb bath. The collision rate $\Gamma_\mathrm{coll}$ is calculated as \cite{Schmidt.2019b}
    \begin{equation}
        \Gamma_\mathrm{coll} = \langle \sigma_\mathrm{el} \ v_\mathrm{coll} \rangle n(\vec{r}_\mathrm{Cs}),
    \end{equation}
    where $\sigma_\mathrm{el}$ is the elastic scattering cross section, obtained from coupled-channel calculations based on the molecular interaction potentials of Rb–Cs, taking into account the internal atomic structure of both species \cite{Schmidt.2019}.
    In our velocity regime, inelastic collisions are negligible.
    $\langle \sigma_\mathrm{el} \ v_\mathrm{coll} \rangle$ denotes the elastic rate coefficient obtained by averaging $\sigma_\mathrm{el} \ v_\mathrm{coll}$ over the relative-velocity distribution. Here, $v_\mathrm{coll}$ denotes the relative collision velocity between the Cs and Rb atoms, and $n(\vec{r}_\mathrm{Cs})$ is the local Rb bath density at the position of the Cs atom.
    Since the bath density is position-dependent, the collision rate throughout transport varies with the position of the Cs atom and, therefore, with time. We can estimate the average number of collisions that one Cs atom undergoes during transport through the Rb bath via the path-averaged collision rate $\bar{\Gamma}_\mathrm{coll}=1/t_\mathrm{coll}\int^{t_\mathrm{coll}}_0 \Gamma_\mathrm{coll}(t) \mathrm{d}t$: 
    \begin{equation}
        \langle N_\mathrm{coll}\rangle=\bar{\Gamma}_\mathrm{coll} \ t_\mathrm{coll}(v),
    \end{equation}
    where $t_\mathrm{coll}(v)$ denotes the time Cs atoms spend traversing the Rb bath. The bath extension $L_\mathrm{bath}$ and the transport velocity determine the interaction time. 
    To model the collision velocity, we assume a Maxwell–Boltzmann distribution which combines the lattice velocity of the Cs atoms with the thermal velocity distribution of the Rb bath. In this approximation, the Cs velocity is taken to be equal to the lattice velocity, while the Rb velocities $v_\mathrm{Rb}$ follow the thermal distribution set by the bath temperature. The relative velocity is given by $\vec{v}_\mathrm{rel}=\vec{v}_\mathrm{Rb}-\vec{v}_\mathrm{lat}$ and its magnitude by $v=|\vec{v}_\mathrm{rel}|$. Integrating the shifted three-dimensional Maxwell-Boltzmann distribution over the solid angle $f(v)= v^2 \int f_\mathrm{rel}(\vec{v})d\Omega$ yields \cite{Xie.2023}
    \begin{equation}
        f(v)=\sqrt{\frac{2}{\pi}}\frac{v}{\sigma v_\mathrm{lat}}\sinh\left( \frac{vv_\mathrm{lat}}{\sigma^2} \right) \exp \left( -\frac{v^2+v_\mathrm{lat}^2}{2\sigma^2} \right)
    \end{equation}
    with $\sigma = \sqrt{\frac{k_\mathrm{B}T_\mathrm{Rb}}{m_\mathrm{Rb}}}$.
    In \cref{fig:n-coll-density-temperature-grid}, we show the expected number of collisions for an uninterrupted Cs trajectory through the entire Rb bath at a lattice velocity of $v_\mathrm{lat}=\SI{7}{\centi\meter\per\second}$. This corresponds to the maximum number of collisions that can be accumulated during transport, since any collision-induced interruption of the trajectory reduces the interaction time with the bath. The number of collisions increases for hotter bath temperatures and for denser baths. We see the same qualitative behavior as in \cref{fig:through_colormeshes} a)-c): colder bath temperatures can decrease the number of collisions and therefore compensate for higher bath densities. This, in turn, leads to increased transport efficiency.
    
\begin{figure*}[t]
    \centering

    \begin{subfigure}[t]{0.49\textwidth}
        \vspace{4pt}
        \centering
        \caption{}
        \includegraphics%
            {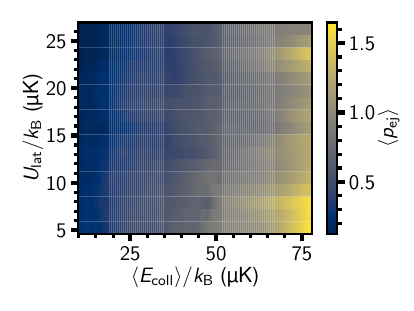}
        \label{fig:eject_prob_mean}
    \end{subfigure}%
    \hfill
    \begin{subfigure}[t]{0.49\textwidth}
        \vspace{4pt}
        \centering
        \caption{}
        \includegraphics%
            {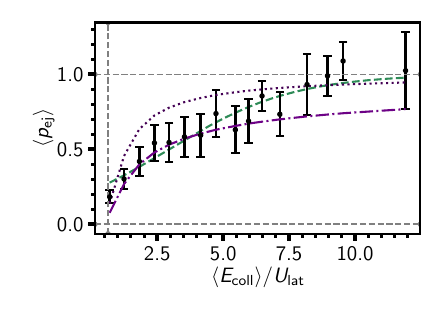}
        \label{fig:ejec_prob_vs_E/U}
    \end{subfigure}
	   \caption[Ejection probability as a function of collision energy]{Ejection probability as a function of collision energy and lattice depth. a) Mean ejection probability $\langle p_\mathrm{ej}\rangle$ as a function of $U_\mathrm{lat}/k_\mathrm{B}$ and $\langle E_\mathrm{coll}\rangle$. b) Mean ejection probability as a function of the dimensionless ratio $\langle E_\mathrm{coll}\rangle /U_\mathrm{lat}$. The error bars denote the standard deviation of the contributing data points. For a monoenergetic collision and an initially stationary Cs atom, the vertical dashed line at $E_{\rm coll}/U_{\rm lat}=1/1.58$ gives minimum collision energy for which direct ejection is kinematically possible and serves as a kinematic reference. The experiment involves finite collision-energy and initial motional distributions as well as subsequent dynamics in the moving lattice potential. The green dashed line represents a phenomenological logistic fit curve describing the crossover from low to high ejection probability. The dotted line indicates the graph of \cref{eq:p_3d}, whereas the dash-dotted line shows \cref{eq:p_z}.}\label{fig:ejec_prob}
    \end{figure*}

    \Cref{fig:ncoll_vs_vlat} shows the estimate for the maximum number of collisions along an uninterrupted Cs trajectory through the Rb bath as a function of the lattice velocity, assuming transport through the whole Rb bath. We find $ N_\mathrm{coll} \sim 0.1 - 12.0$, indicating that individual collision events play a decisive role in determining whether a Cs atom remains trapped in the lattice or becomes localized in the Rb bath. For higher velocities, the number of collisions decreases rapidly. The reduced scattering rate, combined with the shorter interaction time, leads to fewer collisions and thus higher transport efficiency.
    The pointwise relative transport efficiency is given by
    \begin{equation}
        \eta_\mathrm{rel}(v_\mathrm{lat}, U_\mathrm{lat}) = \frac{\eta_\mathrm{t}(v_\mathrm{lat}, U_\mathrm{lat})}{\eta_\mathrm{t, Cs}(v_\mathrm{lat}, U_\mathrm{lat})}
        \label{eq:pointwise_eff}
    \end{equation}
    with the transport efficiency $\eta_\mathrm{t, Cs}$ of the Cs atoms in the lattice without any influence from the Rb bath.
    To relate the mean number of collisions to the relative transport efficiency, we model the collisions as a Poisson process with mean collision number $\langle N_\mathrm{coll} \rangle$. The probability that the impurity undergoes $k$ collisions during interaction with the bath is given by 
    \begin{equation}\label{eq:poisson}
        P(N_\mathrm{coll}=k) = \frac{\langle N_\mathrm{coll} \rangle^k}{k!} e^{-\langle N_\mathrm{coll} \rangle},
    \end{equation}
    and hence the probability for $k=0$ collisions is $P(0)=\exp\left(-\langle N_\mathrm{coll} \rangle\right)$. We further assume that each collision independently ejects the atom
from the lattice with probability $p_\mathrm{ej}$.
    The probability that an atom remains trapped in the lattice and therefore is transported despite undergoing $k$ collisions is 
    \begin{equation}\label{eq:survivial_prob}
        P(\mathrm{survive}|N_\mathrm{coll}=k)=(1-p_\mathrm{ej})^k.
    \end{equation}
    We identify the survival probability associated with Cs–Rb collisions with the relative transport efficiency. Averaging over all possible collision numbers therefore gives
    \begin{align}
        \eta_\mathrm{rel}
        &=
        \sum_{k=0}^{\infty}
        P(N_\mathrm{coll}=k)
        P(\mathrm{survive}\mid N_\mathrm{coll}=k)
        \nonumber\\
        &=
        \sum_{k=0}^{\infty}
        \frac{\langle N_\mathrm{coll} \rangle^k}{k!}
        (1-p_\mathrm{ej})^k \exp\left(-\langle N_\mathrm{coll} \rangle\right)
        \nonumber\\
        &=
        \exp\left(-p_\mathrm{ej}\langle N_\mathrm{coll}\rangle\right).
        \label{eq:effective_ejection_rate}
    \end{align} 
    Consequently, the relative transport loss $1-\eta_\mathrm{rel}$ is given by 
    \begin{equation}
        1-\eta_\mathrm{rel}=1-\exp(-p_\mathrm{ej} \langle N_\mathrm{coll}\rangle).
        \label{eq:effective_ejection_rate_loss}
    \end{equation}
    In the limit $p_\mathrm{ej}\langle N_\mathrm{coll}\rangle \ll 1$, this expression reduces to
    $1-\eta_\mathrm{rel} \simeq p_\mathrm{ej}\langle N_\mathrm{coll}\rangle$.\\
    Even though the calculated collision numbers remain of order unity, an increasing mean number of collisions tends to coincide with larger transport losses.
    Conversion of \cref{eq:effective_ejection_rate} based on the mean ejection probability therefore yields:
    \begin{equation}
        p_\mathrm{ej}(v_\mathrm{lat},U_\mathrm{lat}; n_\mathrm{Rb},T_\mathrm{Rb}) = - \frac{\ln[\eta_\mathrm{rel}(v_\mathrm{lat},U_\mathrm{lat}; n_\mathrm{Rb},T_\mathrm{Rb})]}{\langle N_\mathrm{coll}(v_\mathrm{lat}; n_\mathrm{Rb},T_\mathrm{Rb})\rangle}.
        \label{eq:ejection_probability}
    \end{equation}
    Equation (\ref{eq:ejection_probability}) converts the ensemble transport measurement into a single-collision observable: $p_{\rm ej}$ is the conditional probability that one collision produces a detectable transport failure.
        We can determine the mean relative kinetic energy as 
    \begin{equation}
        \langle E_\mathrm{coll} \rangle = \frac{\int 1/2\mu v^2 \,\sigma_\mathrm{el}(v)  v  f(v)\mathrm{d}v}{\int \sigma_\mathrm{el}(v)  v  f(v)\mathrm{d}v},
        \label{eq:collision_energy}
    \end{equation}
    where $\mu$ is the reduced mass of Cs and Rb. 
    Combining the measured transport efficiency with the calculated collision number therefore allows us to extract $p_{\rm ej}$ for each set of collision and lattice parameters.
    By averaging points of similar collisional energy and lattice depth, the mean ejection probability $\langle p_\mathrm{ej}\rangle$ can thus be determined, which is shown in \cref{fig:eject_prob_mean}. 
    The ejection probability generally increases with increasing collision energy and decreasing lattice depth, indicating that the relative magnitude of these two energy scales governs the transport stability. 
    At high collision energies, the mean collision number becomes small, making the inversion in \cref{eq:ejection_probability} prone to fluctuations in $\eta_\mathrm{rel}$.  
    Consequently, some point estimates of $\langle p_\mathrm{ej} \rangle$ exceed the physical upper bound of unity. 
    We therefore interpret these values as being compatible with $\langle p_\mathrm{ej} \rangle \simeq 1$ within the experimental uncertainty rather than as probabilities greater than one.     
    Scanning $v_{\rm lat}$ changes the collision-energy distribution, whereas scanning $U_{\rm lat}$ changes the detector threshold. 
    Their ratio therefore provides a dimensionless measure of collision energy relative to the detection scale.
    To make this dependence explicit, \cref{fig:ejec_prob_vs_E/U} shows the same data as a function of the dimensionless ratio $\langle E_\mathrm{coll}\rangle /U_\mathrm{lat}$. 
    For $\langle E_\mathrm{coll}\rangle /U_\mathrm{lat} \ll 1$, the ejection probability decreases toward zero, whereas it approaches unity at large energy ratios. 
    We describe this crossover phenomenologically using a logistic function, shown by the green line.
    Viewed as a threshold measurement, the dependence of $p_{\rm ej}$ on $U_{\rm lat}$ probes the cumulative distribution of collision-induced energy transfer and therefore provides a route toward threshold spectroscopy of individual ultralow-energy collision events.

    Taking into account the corresponding collision energies, determined by lattice velocity and thermal motion of the Rb atoms, we obtain values in the range of \mbox{$\langle E_\mathrm{coll}\rangle/k_\mathrm{B}=$\ \SIrange{1}{130}{\micro\kelvin}.} 
    To obtain a simple kinematic reference for the crossover shown in \cref{fig:ejec_prob_vs_E/U}, we consider an elastic collision at fixed collision energy and approximate the Cs atom as initially stationary in the lattice frame. The energy transferred to Cs in an elastic collision with center-of-mass scattering angle $\theta$ becomes \cite{Lim.2015}
    \begin{equation}
        \Delta E_\mathrm{Cs} = \left( \frac{2m_\mathrm{Rb}}{m_\mathrm{Rb}+m_\mathrm{Cs}} \right ) (1-\cos(\theta)) E_\mathrm{coll}.\label{eq:energy_transfer}
    \end{equation}
    The transferred energy is maximal for backscattering, where $\theta=\pi$, such that up to  $\Delta E_\mathrm{Cs, max}=\frac{4 m_\mathrm{Rb}}{m_\mathrm{Rb}+m_\mathrm{Cs}} E_\mathrm{coll} \approx 1.58 E_\mathrm{coll}$ can be transferred to the Cs atom \cite{LandauLifshitz}. 
    Thus, the energy associated with a single elastic collision can in principle significantly exceed the lattice depth, especially at higher lattice velocities. Direct ejection by a single collision is therefore kinematically possible if $E_\mathrm{coll}/U_\mathrm{lat}\gtrsim1/1.58$.
    The collider hence operates as a threshold-sensitive detector of single-collision energy transfer rather than as a conventional force probe. The corresponding monoenergetic threshold is indicated by the vertical dashed line in \cref{fig:ejec_prob_vs_E/U}.
    Importantly, this bound corresponds to the maximum possible energy transfer in a backscattering event and therefore does not set the expected crossover scale.  
    For isotropic $s$-wave scattering, $\cos (\theta)$ is uniformly distributed over the range [-1,1]. Since the transferred energy in \cref{eq:energy_transfer} depends linearly on $1-\cos (\theta)$, $\Delta E_\mathrm{Cs}$ is uniformly distributed between 0 and $\simeq1.58E_\mathrm{coll}$. 
    In the least restrictive threshold model, we assume that ejection occurs whenever the total energy transferred to the Cs atom exceeds the lattice depth, $\Delta E_\mathrm{Cs}>U_\mathrm{lat}$.  The corresponding ejection probability is 
    \begin{align}
        p^{(E)}_\mathrm{ej}=P(\Delta E_\mathrm{Cs}>U_\mathrm{lat}) & = \int_{U_\mathrm{lat}}^{1.58 E_\mathrm{coll}}\frac{\mathrm{d}(\Delta E_\mathrm{Cs})}{1.58 E_\mathrm{coll}}\\ &= 1-\frac{U_\mathrm{lat}}{1.58 E_\mathrm{coll}}
        \label{eq:p_3d}
    \end{align}
    for $E_\mathrm{coll}/U_\mathrm{lat}>1/1.58$, and zero otherwise. This idealized total-energy limit is indicated by the dotted line in \cref{fig:ejec_prob_vs_E/U}.
    Since transport in our experiment occurs predominantly along the $z-$direction, we additionally consider the longitudinal component of the transferred energy.
    For an incident relative velocity along $z$,
    \begin{equation}
        \Delta E_{\mathrm{Cs},z}=\Delta E_{\mathrm{Cs, max}} x^2
    \end{equation}
    with $x=\frac{1-\cos(\theta)}{2}$. Since for isotropic $s$-wave scattering, $x$ is uniformly distributed on [0,1] and
    \begin{equation}
        P(\Delta E_{\mathrm{Cs},z}>U_\mathrm{lat}) = P\left(x>\sqrt{\frac{U_\mathrm{lat}}{1.58 E_\mathrm{coll}}}\right),
    \end{equation}
    the longitudinal-energy limit is given by
    \begin{equation}
        p^z_\mathrm{ej} = 1-\sqrt{\frac{U_\mathrm{lat}}{1.58 E_\mathrm{coll}}}.
        \label{eq:p_z}
    \end{equation}
    The dash-dotted line in \cref{fig:ejec_prob_vs_E/U} shows the longitudinal-energy limit.
    The experimentally measured crossover lies between these two idealized kinematic limits over much of the experimentally relevant range. Because both of these limits neglect the finite collision-energy distribution, the initial Cs motion, and the subsequent dynamics in the moving periodic potential, we use them as parameter-free benchmarks for our data. 
    A phenomenological fit yields a characteristic crossover at $E_{\rm coll}/U_{\rm lat} = 2.95\pm 0.27$ for $p_{ej}=0.5$, providing an empirical energy scale for collision-induced transport disruption.
    The observed crossover occurs on the energy scale expected for single-collision disruption of lattice transport.

\section{Conclusion}\label{conclusion}
    In conclusion, we studied the controlled transport of single Cs atoms in a shallow optical lattice through an ultracold thermal bath of Rb atoms. 
    This regime is usually hidden for conventional transport experiments, where dissipation is typically represented as a continuous drag force instead of individual collisions. 
    The shallow lattice acts as an energy-sensitive discriminator that converts collisional energy transfer into a binary transport observable, i.e. the impurity either remains successfully transported or its transport is disrupted.
    By varying the bath density and temperature, along with the lattice velocity and depth, we can independently tune the experimental parameters that govern the mean number of collisions, their characteristic energy, and the threshold for disrupting transport. 
    The observed losses increase with the calculated collision number. 
    They are also consistent with a Poisson survival model in which each collision has an effective probability of ejecting the Cs atom from the lattice.
    This ejection probability increases as the mean collision energy becomes comparable to or exceeds the lattice depth, supporting the interpretation that individual elastic collisions can interrupt transport.
    This allows us to extract a single-collision ejection probability and to directly resolve how the microscopic transport response depends on the ratio of collision energy to confinement energy.
    The experiment thereby establishes an atomic-billiard platform that connects controlled ultralow-energy scattering with dissipative transport. 
    Its independently tunable collision energy and detection threshold provide a route toward threshold-resolved collision spectroscopy and, in future work, toward probing energy-transfer statistics in degenerate or nonequilibrium baths.

\begin{acknowledgments}
The authors acknowledge Alexander Guthmann for providing
the scattering cross-section data.
We thank Jonas Gutsche and Nessa Baum for insightful
comments and carefully reading the manuscript.\\
This work was supported by the Deutsche
Forschungsgemeinschaft (DFG, German Research Foundation)
through Collaborative Research Center SFB/TR185
(Project No. 277625399).\\
S.B. acknowledges funding from the Studienstiftung
des deutschen Volkes. A.W. acknowledges funding from the European Research
Council (ERC) under the European Union's Horizon Europe
research and innovation programme
(ERC Advanced Grant No. 101200776).

\end{acknowledgments}

\hspace{3pt}

\section*{Data Availability Statement}
The data that support the findings of this study are openly available at the following URL/DOI: \url{https://doi.org/10.5281/zenodo.23671017}.

    \bibliographystyle{apsrev4-2}
	\bibliography{bibliography_paper}

\newpage
\appendix

\section{Detailed Cs transport scheme}\label{sec:lat_transport_details}
    In this section, we describe the Cs transport protocol in detail, including the definition of spatial regions of interest and the kinematic scheme used to conduct transport efficiency measurements.\\
    To quantify the transport efficiency, we define three regions of interest along the $z$-axis (transport axis): the preparation region for Cs atoms, the Rb region, where the Rb bath is located, and the target region.\\
    The position of the preparation region is given by the location of the crossed optical dipole trap used to confine the Cs atoms. For the subsequent analysis, however, we assign to this region an effective width of \SI{140}{\micro\meter}, which is substantially larger than the initial spatial extent of the Cs atom distribution of about \SI{22+-4}{\micro\meter}. This choice is justified by the observation that atoms that are not successfully transported typically leave the confinement of the optical lattice and consequently exhibit pronounced spatial spreading.
    
    For the transport efficiency measurements presented in \cref{sec:meas_cs} and \cref{meas_through}, we analyze the transport dynamics and, in turn, the final positions of the atoms between the Cs preparation region and the target region. \\
    We can control the lattice motion by changing the detuning $\Delta \omega_L= \omega_{L1}-\omega_{L2}$ between the two laser beams of frequencies $\omega_{L1}$ and $\omega_{L2}$. In this way, the velocity $v$ is given by
    \begin{equation}
        v=\frac{\Delta \omega_L}{2k}
    \end{equation}
    with wavevector $k=2\pi / \lambda_L$. To transport the atoms, we detune the lattice beams for a time $t_\mathrm{transp}$, such that
    \begin{equation}
        t_\mathrm{transp}=\sqrt{\frac{9d}{2a}},
    \end{equation}
    where $d$ is the targeted transport distance. This time is given by our trapezoidal transport profile. The transport sequence consists of three intervals of equal duration $\Delta t$: acceleration, constant-velocity motion, and deceleration. Thus, $t_\mathrm{transp}=3\Delta t$.\\
    The plateau velocity is
    \begin{equation}
        v=a\Delta t = \sqrt{\frac{d a}{2}}.
    \end{equation}
    We accelerate the Cs atoms in the first part of the transport such that the atoms have a constant velocity when they are inside the densest part of the Rb bath. After leaving the Rb bath, we decelerate the atoms to zero lattice velocity to allow imaging of their positions along the $z$ axis. Due to technical constraints, we cannot determine the positions of the Cs atoms perpendicular to the $z$ axis.\\

\section{Lattice influence on the Rb bath}\label{sec:rb_influenced}
    In this section, we examine the influence of the transport lattice on the Rb clouds. We apply the same lattice-power and transport sequence as in the transport measurements but without inserting the Cs atoms into the transport lattice. After the lattice transport, we image the Rb clouds. \\
        \begin{figure*}[bth]
    	   \includegraphics{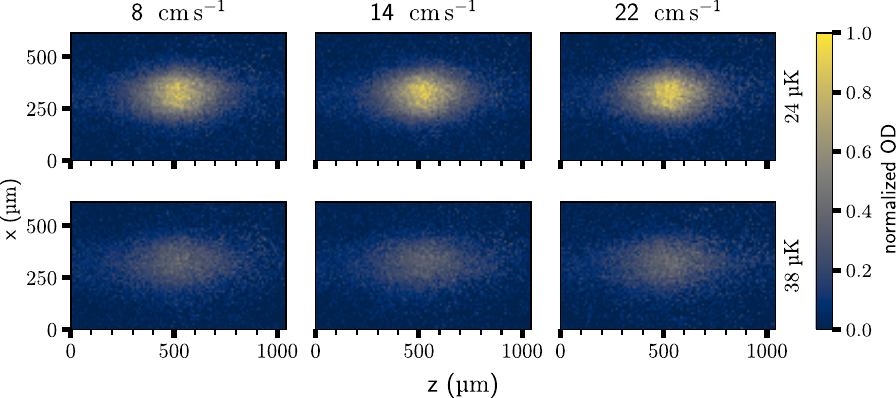}
    	   \caption[Lattice influence on the Rb clouds]{Influence of the transport lattice on the Rb cloud for different lattice depths and velocities. Each column shows results for one lattice velocity, while the rows represent two different lattice depths (given in units of temperature). The column density is reduced as the lattice depth increases. For a lattice depth of $U_\mathrm{lat}/k_\mathrm{B}=\SI{38}{\micro\kelvin}$, the cloud densities are visibly reduced, and their profile is elongated in the $z$-direction. 
           }\label{fig:Rb_influence}
        \end{figure*}
    We measure the optical density of the Rb clouds for different lattice depths and accelerations; see \cref{fig:Rb_influence}. Each column shows the absorption images for one of the three exemplary velocities: 8, 14, and \SI{22}{\centi\meter\per\second}, while the rows show the results for lattice depths of 24 and \SI{38}{\micro\kelvin}, quoted in temperature units. The bath density decreases significantly at larger lattice depths, preventing the reliable production of dense Rb clouds. However, the peak optical density varies visibly for the different lattice velocities. Because the transport distance is fixed, changing the transport velocity also changes the total time for which the lattice is applied.
    This time ranges from \SI{4.2}{\milli\second} at the highest lattice acceleration to \SI{23.9}{\milli\second} at the lowest. % 130 m/s^2 vs 4 m/s^2
    The influence on the Rb cloud increases with lattice depth. For $U_\mathrm{lat}/k_\mathrm{B}=$\SI{38}{\micro\kelvin}, the Rb cloud is heated and expands sufficiently to preclude transport measurements through a dense bath. Therefore, we restrict our measurements to $U_\mathrm{lat}/k_\mathrm{B}$ well below \SI{30}{\micro\kelvin}. 

\section{Cs transport efficiencies for low lattice accelerations in detail}\label{sec:cs_only_low_acc}
\begin{figure*}[hbt]
    \begin{subfigure}{0.49\textwidth}
              \caption{}
              \includegraphics{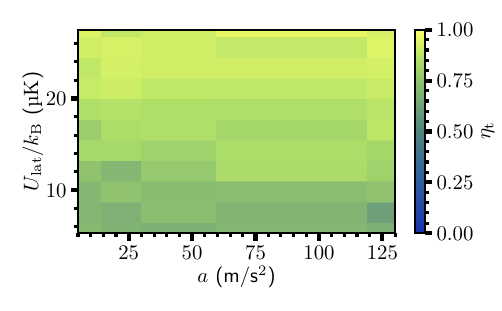}
        \end{subfigure}
        \hfill
        \begin{subfigure}{0.49\textwidth}
              \caption{}
	       \includegraphics{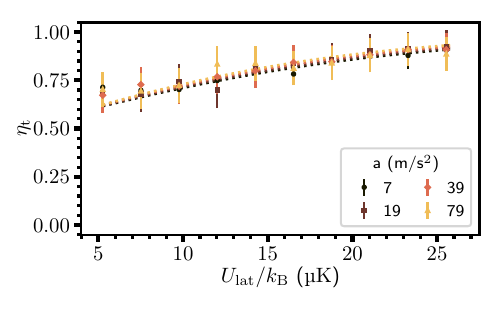}
        \end{subfigure}
	   \caption[Cs transport measurement efficiencies for low accelerations and lattice depths]{Transport efficiencies of Cs atoms in the optical lattice for low lattice accelerations and lattice depths. The parameter range matches that used for the measurements in \cref{fig:through_colormeshes}. Panel (b) shows selected accelerations from the full parameter range shown in panel a). The dashed lines in panel (b) represent fits with a logistic function.}\label{fig:Cs_transport_small_params}
    \end{figure*}

    To isolate collision-induced changes in transport efficiency, we use relatively low lattice accelerations. Otherwise, the Cs atoms could not be reliably transported within the lattice, even without collisions with the Rb bath. To match these constraints, we chose initial lattice accelerations between 8 and \SI{130}{\meter\per\second\squared}, corresponding to velocities from \SI{0.045}{\meter\per\second} to \SI{0.182}{\meter\per\second} during the interaction. The corresponding transport efficiencies for these parameters are shown in \cref{fig:Cs_transport_small_params}.
    \par
    Transport efficiencies across the entire investigated parameter range remain well above 50\% and increase monotonically with lattice depth. In contrast, varying the lattice acceleration does not yield statistically significant changes in this parameter regime, with all measured values agreeing within their respective error bars.
    
\section{Mean free path}\label{sec:mfp}
\begin{figure}[t]
\centering
        \includegraphics{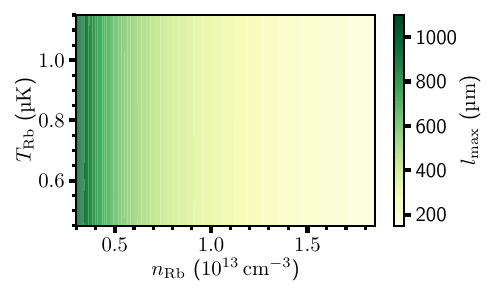}
	   \caption[Maximum mean free path of Cs atoms in the Rb bath]{Maximum mean free path of the Cs atoms in the Rb bath optimized for Cs velocities from \SIrange{4}{18}{\centi\meter\per\second}.}\label{fig:mfp_max}
    \end{figure}

    Calculating the mean free path of Cs atoms in the Rb bath supports our finding that few collisions determine transport behavior. 
    To compare the distance traveled between two collisions with the spatial extent of the Rb bath, we define the mean free path of the Cs atoms as \cite{Chapman.1998}
    \begin{equation}
        l(v_\mathrm{lat}) = \frac{v_\mathrm{lat}}{\Gamma_\mathrm{coll}(v_\mathrm{lat})}=\frac{v_\mathrm{lat}}{n_\mathrm{Rb}\langle\sigma_\mathrm{el} v_\mathrm{rel}\rangle}.
    \end{equation}\\
        \begin{table}[t]
        \centering
        \small
        \setlength{\tabcolsep}{3pt}
        \caption[Mean free path for the bath parameters in \cref{fig:eta-transp-through}]{Mean free paths for the bath parameters shown in \cref{fig:eta-transp-through}.}
        \label{tab:mfp}
        \begin{tabular}{c|c|c|c|c}
             $T_\mathrm{Rb}$& $n_\mathrm{peak, Rb}$ & $l_\mathrm{max}$ & $l_{v=\SI{8}{\centi\meter\per\second}}$& $l_\mathrm{min}$ \\
             $(\SI{}{\nano\kelvin})$& (\SI{}{\per\centi\meter\cubed}) & (\SI{}{\micro\meter}) & (\SI{}{\micro\meter})& (\SI{}{\micro\meter})
             \Bstrut\\ \hline
             \Tstrut
             504 & \SI{1.80e13}{} & \SI{173.4+-6.3}{} & \SI{19.6+-0.2}{} & \SI{6.3+-0.1}{}\\
             549 & \SI{1.17e13}{} & \SI{266.4+-9.6}{} & \SI{30.2+-0.4}{} & \SI{9.6+-0.1}{}\\
             793 & \SI{1.07e13}{} & \SI{292.8+-10.7}{} & \SI{33.3+-0.4}{} & \SI{10.7+-0.1}{}\\
             865 & \SI{1.81e13}{} & \SI{172.3+-6.3}{} & \SI{19.6+-0.2}{} & \SI{6.3+-0.1}{}\\
             887 & \SI{1.45e13}{} & \SI{215.3+-7.9}{} & \SI{24.4+-0.2}{} & \SI{7.9+-0.1}{}\\
             918 & \SI{1.15e13}{} & \SI{270.6+-9.9}{} & \SI{30.8+-0.2}{} & \SI{9.9+-0.1}{}\\
             931 & \SI{1.01e13}{} & \SI{308.2+-11.3}{} & \SI{35.0+-0.3}{} & \SI{11.3+-0.1}{}\\
             956 & \SI{0.60e13}{} & \SI{524.7+-19.3}{} & \SI{59.7+-0.8}{} & \SI{19.3+-0.3}{}\\
             1054 & \SI{0.30e13}{} & \SI{1035.2+-38.2}{} & \SI{117.8+-1.8}{} & \SI{38.2+-0.6}{}\\
        \end{tabular}         
        \end{table}
    This definition accounts for thermal motion of the individual Rb atoms and the velocity dependence of the elastic cross section.
    We expect a significant dependence on the bath density and the collision velocity, since this heavily influences the scattering cross-section.\\
    The lowest mean free path can be reliably found at \SI{4}{\centi\meter\per\second}, while the maximum corresponds to the highest velocity of \SI{18}{\centi\meter\per\second}. \Cref{tab:mfp} summarizes the bath parameters resulting from \cref{fig:eta-transp-through} for different velocities. At a velocity of \SI{8}{\centi\meter\per\second}, the mean free path is already comparable to the bath's spatial extent, supporting the interpretation that transport occurs in the few-collision regime and that single collisions determine the transport outcome. \Cref{fig:mfp_max} shows the maximum mean free path over the entire range of bath parameters. While the bath temperature has only a minor effect, the bath density is the dominant factor influencing the mean free path of Cs atoms.
    
\end{document}